\documentclass[journal]{IEEEtran}

\IEEEoverridecommandlockouts

\usepackage{cite}
\usepackage{amsmath}
\usepackage{bm}
\usepackage{subcaption}
\usepackage{amssymb,amsmath}
\usepackage[mathscr]{euscript}
\usepackage{graphicx}
\usepackage{textcomp}
\usepackage{xcolor}
\usepackage{multirow}
\usepackage{booktabs}
\usepackage{algpseudocode}
\usepackage[linesnumbered,ruled,vlined]{algorithm2e}

\def\BibTeX{{\rm B\kern-.05em{\sc i\kern-.025em b}\kern-.08em
    T\kern-.1667em\lower.7ex\hbox{E}\kern-.125emX}}
\begin{document}
\bstctlcite{IEEEexample:BSTcontrol}
\title{Efficient Resource Management for Secure and Low-Latency O-RAN Communication}
\author{\IEEEauthorblockN{Zaineh~Abughazzah${}^{1}$, Emna~Baccour${}^2$, Ahmed~Refaey${}^3$, 
Amr~Mohamed${}^1$ and Mounir~Hamdi${}^2$} \\
\IEEEauthorblockA{
${}^1$ College of Engineering, Qatar University, Doha, Qatar.\\
${}^2$ College of Science and Engineering, Hamad Bin Khalifa University, Qatar Foundation, Doha, Qatar.\\
${}^3$ School of Engineering, 
University of Guelph, Guelph, Ontario, Canada.\\
}
}
\maketitle

\begin{abstract}
Open Radio Access Networks (O-RAN) are transforming telecommunications by shifting from centralized to distributed architectures, promoting flexibility, interoperability, and innovation through open interfaces and multi-vendor environments. However, O-RAN’s reliance on cloud-based architecture and enhanced observability introduces significant security and resource management challenges.
Efficient resource management is crucial for secure and reliable communication in O-RAN, within the resource-constrained environment and heterogeneity of requirements, where multiple User Equipment (UE) and O-RAN Radio Units (O-RUs) coexist. This paper develops a framework to manage these aspects, ensuring each O-RU is associated with UEs based on their communication channel qualities and computational resources, and selecting appropriate encryption algorithms to safeguard data confidentiality, integrity, and authentication. A Multi-objective Optimization Problem (MOP) is formulated to minimize latency and maximize security within resource constraints. Different approaches are proposed to relax the complexity of the problem and achieve near-optimal performance, facilitating trade-offs between latency, security, and solution complexity. Simulation results demonstrate that the proposed approaches are close enough to the optimal solution, proving that our approach is both effective and efficient.
\end{abstract}


\section{Introduction} \label{sec:introduction}
In the evolving landscape of telecommunications, O-RAN represents a paradigm shift towards flexible and interoperable network architectures, introducing new security challenges, particularly in data protection \cite{s24031038}. Encryption ensures data confidentiality, integrity, and authentication, protecting sensitive information from unauthorized access and tampering during transmission in addition to verifying the identities of communicating parties and ensuring that data is exchanged between legitimate UEs and O-RUs. The unique aspects of O-RAN, such as open interfaces, a multi-vendor environment, and reliance on cloud-based architecture, necessitate adaptable encryption protocols tailored to its programmable and flexible nature. Enhanced observability and control capabilities in O-RAN, while beneficial for network management, require strong encryption to prevent exploitation by malicious actors \cite{UnderstandORAN}. For instance, if a malicious actor impersonates a Service Management and Orchestrator (SMO) or a Near-real-time RAN Intelligent Controller (Near-RT RIC) to access resources, or if legitimate users unintentionally disclose data content, it can lead to confidential data leakage, compromising the security indicator’s confidentiality. Thus, encryption in O-RAN not only enhances security but also builds user trust and ensures compliance with regulatory standards. In a scenario of a military operation, a mobile command unit relies on O-RUs for secure, flexible, low-latency communication. These O-RUs ensure coverage and connectivity for devices like radios, UAVs, and sensors. Security is crucial to protect sensitive information during the mission. Robust encryption protocols are essential, along with power efficiency due to limited power sources. Minimizing latency is critical for real-time communication and rapid response. Ensuring secure communications, efficient power usage, and low latency is vital for mission success and safety.

Many studies in the literature have explored various approaches to enhance communication in wireless networks, focusing on aspects such as latency, security, and reliability.
Authors in \cite{4358705} propose a framework that optimizes encryption parameters based on channel conditions to quantify the trade-off between security and throughput by formulating optimization problems to maximize throughput while maintaining security constraints. Works by \cite{9605044, 9838678} aim to minimize energy consumption while providing delay guarantees to users in O-RAN networks. In \cite{9605044}, the authors propose a joint optimization for resource allocation and distributed units’ selection showing improved energy efficiency compared to disjoint approaches. While authors in \cite{9838678} focus on computation offloading in Internet of Things (IoT) systems with experimental results showing significant reduction.

To the best of our knowledge, existing research on O-RAN resource management often neglects the combined consideration of latency and security during data offloading, as well as user and device heterogeneity. Our framework addresses these gaps by adapting security measures to available resources, ensuring optimal performance. Additionally, balancing the trade-off between maximizing security and minimizing latency is essential. This paper aims to address these objectives concurrently, with a focus on efficient resource utilization.

The main contributions of this paper are as follows:
 \begin{itemize}
    \item We profile the computation complexities of different encryption algorithms to understand the trade-off between security and latency.
     \item We formulate a MOP for association and security level selection, that aims to minimize the total latency in the system and maximize security.
     \item Being non-convex and NP-hard to solve, we propose two different approaches that employ sub-optimal solutions to the relaxed and convexified formulated MOP.
     \item The performance of both proposed approaches is evaluated, analyzed and compared to a benchmark.
 \end{itemize}
The rest of this paper is structured as follows: Section \ref{sec:system_model} introduces the system model, Section \ref{sec:problem_formulation} outlines the formulated MOP, Section \ref{sec:solution} describes the proposed solution approaches, Section \ref{sec:results} shows the simulation results, and Section \ref{sec:conclusion} provides the concluding remarks of our work.

\section{System model} \label{sec:system_model}

\begin{figure}[ht!]
    \centering
    \fbox{\includegraphics[width=1\linewidth, height=0.26\textheight]{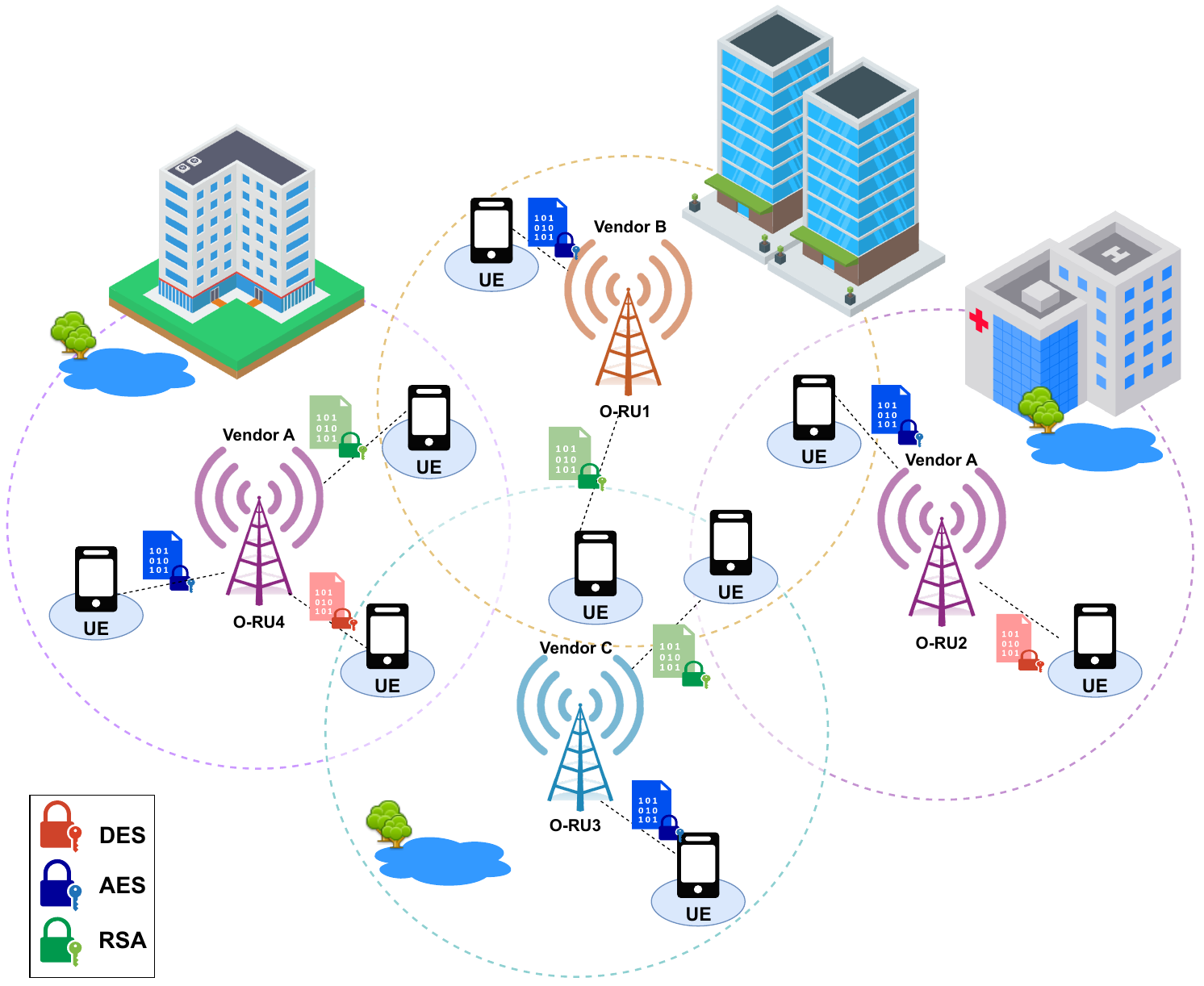}}
    
    \caption{System Model}
    \label{fig:sysmodel}
\end{figure}

As shown in Figure \ref{fig:sysmodel}, we consider a 5G network consisting of a set of O-RAN Radio Units (O-RUs) located within an O-RAN denoted as $\mathcal{J} = \{1,2,...,J\}$ Where $J$ is the total number of O-RUs provided by different vendors. Let $\mathcal{P}_{j}$ denote the processor’s clock speed of the O-RU. Each O-RU has specific security requirement denoted as $W_j$, and Resource Blocks (RBs), which is a basic unit of resource allocation in wireless communication systems that are used for data transmission, allowing multiple UEs to send and receive data simultaneously within the same frequency band, denoted as $M_j$. Additionally, we have a set of mobile UEs denoted as $\mathcal{I} = \{1,2,...,I\}$ where $I$ is the total number of devices. Let $\mathcal{Q}_{i}$ denote the processor's clock speed, $Z_i$ denote battery capacity, and $\Gamma_i$ present its computational power. Moreover, we assume that each UE $i \in \mathcal{I}$ is connected to one O-RU $j \in \mathcal{J}$ at each time step $t \in \mathcal{T}$, where set $\mathcal{T}$ represents discrete time steps, and its length is the total number of time steps considered in the analysis. At each time step $t$, each UE intend to send a data of size $PT_{i,t}$. To ensure optimal connectivity, UEs may switch from one  O-RU to another based on various conditions, including capacity of the UE to respect the security requirements and mobility, which affects bandwidth, and hence the latency. These conditions are evaluated at each time step to determine the most suitable O-RU for each UE. Considering battery capacity, it is crucial to ensure that UEs can achieve and maintain connectivity throughout the duration. This involves managing battery life by optimizing processing and transmission.

\subsection{Encryption algorithms} \label{Encryption_algorithms}
 
When the data is transmitted from the UE to the O-RU, an encryption algorithm is adopted to protect the confidentiality of user data during transmission. In this paper, we assume that key exchange and agreement protocols have been successfully completed, as these are beyond the scope of our current focus.

We consider a list $\mathcal{K}$ of encryption algorithms where each algorithm $ k \in \mathcal{K}$ has a set of possible key sizes denoted as \( D_k \), where the maximum possible key length is equal to $D^k_{\text{max}}$ and the minimum possible key length is equal to $D^k_{\text{min}}$. At each time step during data transmission from each UE to the selected O-RU, a single encryption algorithm is selected. In this paper and without loss of generality, we select three different encryption algorithms with varying key sizes for detailed analysis. The first considered algorithm is the Data Encryption Standard (DES) with a key size of 64 bits. DES is one of the earliest encryption standards, laid the groundwork for modern cryptography and it typically uses a single key size and is considered less secure compared to modern algorithms due to its vulnerability to brute-force attacks \cite{FIPS46-3}. We also consider variants of the Advanced Encryption Standard (AES) algorithm with key sizes of 128, 192, and 256 bits. AES, with its multiple key sizes, offers a higher level of security and efficiency, and has become the current standard for securing sensitive data. Hence, it is widely adopted in various applications\cite{FIPS197-upd1}. Additionally, we consider the Rivest–Shamir–Adleman (RSA) algorithm with key sizes of 1024, 2048, 3072, and 4096 bits. \cite{10.1145/359340.359342}. The selection of these three algorithms for this study is based on their proven effectiveness and widespread use in both literature and real-world applications \cite{article}. 

Hence, we denote the list of possible algorithms by $ k \in \mathcal{K} = \{1,2,3\}$, where:
\begin{footnotesize}
\begin{equation}\label{eq:algorithms}
    k = \left\{\begin{matrix}
\text{$1$, DES algorithm}\\
\text{$2$, AES algorithm}\\
\text{$3$, RSA algorithm}
\end{matrix}\right. 
\end{equation}
\end{footnotesize}

It is worth noting that plaintext is divided into blocks of length $\mathcal{B}_k$ according to the encryption algorithm. Each of these blocks is then encrypted independently using Electronic Code Book (ECB) mode of operation.
This means that each plaintext block of length $\mathcal{B}_k$ is transformed into a ciphertext block of the same length. We choose ECB for simplicity purposes. The block size for each encryption algorithm $ k \in \mathcal{K}$ is as follows:
\begin{footnotesize}
\begin{equation}\label{eq:algorithms2}
   \mathcal{B}_k  = \left\{\begin{matrix}
\text{$\mathcal{B}_{DES}$ = 64}\\
\text{$\mathcal{B}_{AES}$ = 128}\\
\text{$\mathcal{B}_{RSA}$ = $N$}
\end{matrix}\right. 
\end{equation}
\end{footnotesize}
where $N$ is the key length. In this study, we assume that each block is encrypted independently. Subsequently, the entire encrypted data is transmitted. 

In order to encrypt plaintext, each UE adopt an encryption algorithm $k \in \mathcal{K}$ and a secret key. It is worth mentioning that our study includes both symmetric (e.g., DES, AES) and asymmetric (e.g., RSA) encryption methods. Symmetric encryption uses the same key for encryption and decryption, while asymmetric encryption uses a pair of public and private keys. The encryption process transforms the plaintext into ciphertext which is unintelligible without the corresponding decryption key. The ciphertext is sent to the corresponding O-RU. To retrieve the original plaintext, the same encryption algorithm with the corresponding key are used by the O-RU. The size of the data to be transmitted from UE $i \in \mathcal{I}$ at time step $t \in \mathcal{T}$ is denoted as $PT_{i,t}$ (i.e., plaintext) while the size of the data being transmitted ${CT_{i,t}}$ (i.e., ciphertext).
\subsection{Communication and computation models} \label{Communication_computation_models}
\subsubsection{Communication model} 
The communication latency describes the transmission delay of the data being offloaded from the UE $i$ to the O-RU $j$, where it depends on the size of the data being transmitted ${CT_{i,t}}$, and the data transmission rate, $\rho_{i,j,t}$, which is the speed at which digital data is transmitted over a communication channel:  

\begin{equation}
    \tau_{comm}^{i,j,t} = \frac{CT_{i,t}}{\rho_{i,j,t}}
\end{equation}
where $\rho_{i,j,t}$ depends on bandwidth, transmission power, and channel gains of devices, and it is governed by the mobility that dynamically changes over time. For simplicity, we abstracted $\rho_{i,j,t}$ as an indirect function of these factors. Higher $\rho_{i,j,t}$ indicates a shorter distance between a UE and an O-RU, leading to faster data transmission and reduced latency.
 
\subsubsection{Computation model}
Encryption and decryption latencies describe the time delay introduced by the computational overhead required to perform encryption and decryption operations. This latency is affected by multiple factors including the encryption algorithm used, the hardware acceleration available, the efficiency of the software implementation, and the degree of parallelization. 


We denote the computational complexity to encrypt or decrypt one block of data $\mathcal{B}_k$ using encryption algorithm $k \in \mathcal{K}$ by $C_{k}$. As previously described, we consider 3 encryption algorithms: DES, AES, and RSA, and their complexities are computed as follows \cite{zwmm}:

\begin{footnotesize}
\begin{equation}
    C_{k} = \left\{
    \begin{matrix}
    C_{DES_{enc}} = C_{DES_{dec}}= 16 \cdot N_{shift}\\ + R_{DES}( 10 \cdot N_{shift} +  10 \cdot N_{xor})\\
    \\
    C_{AES_{enc}} = 16 \cdot N_{xor} + (R_{AES}-1)( 184 \cdot N_{and}\\ +  136 \cdot N_{or} + 352 \cdot N_{shift})\\ + ( 16 \cdot N_{xor} +  12 \cdot N_{shift} + 12 \cdot N_{or})\\
    \\
    C_{AES_{dec}} = 16 \cdot N_{xor} + (R_{AES}-1)( 644 \cdot N_{and} \\+  500 \cdot N_{or} + 224 \cdot N_{shift}) \\+ ( 16 \cdot N_{xor} +  12 \cdot N_{shift} + 12 \cdot N_{or})\\
    \\
     C_{RSA_{enc}} = C_{RSA_{dec}} = N^2\\
\end{matrix}\right.
\end{equation}
\end{footnotesize}
where $N$ is the key length, and $R_k$ is the number of rounds for the chosen encryption algorithm (i.e., 16 in DES, 10 in AES-128, 12 in AES-192, and 14 in AES-256):

\begin{footnotesize}
\begin{equation}\label{eq:algorithms3}
   R_k  = \left\{\begin{matrix}
\text{$R_{DES} = 16$}\\
\text{$R_{AES} = \frac{N}{32} + 6$}\\
\text{$R_{RSA} = 1$}
\end{matrix}\right. 
\end{equation}
\end{footnotesize}
while $N_{and}$, $N_{or}$, $N_{shift}$, $N_{xor}$ are the number of cycles required to perform one AND operation, one OR operation, one shift operation, and one XOR operation respectively. Based on the Intel x86 instruction set, $N_{and}$ = 1 clock cycle, $N_{or}$ = 1 clock cycle, $N_{shift}$ = 1 clock cycle, while $N_{xor}$ = 3 clock cycles 
\cite{shanley2009x86}.

Accordingly, the latency for encryption depends on the data size $PT_{i,t}$, the computational complexity of the encryption algorithm $C_k$, and the processor’s clock speed of the UE $\mathcal{Q}_{i}$:
\begin{footnotesize}
\begin{equation}
    \tau_{enc}^{i,k,t} =  C_{k} \frac{\lceil \frac{PT_{i,t}}{\mathcal{B}_{k}} \rceil}{\mathcal{Q}_{i}}
\end{equation}
\end{footnotesize}
Similarly, the latency for decryption depends on the number of bits in the encrypted data $CT_{i,t}$, the computational complexity of the decryption $C_k$, and the processor’s clock speed of the O-RU $\mathcal{P}_{j}$: 
\begin{footnotesize}
\begin{equation}
    \tau_{dec}^{j,k,t} = C_{k} \frac{\lceil \frac{CT_{i,t}}{\mathcal{B}_{k}} \rceil}{\mathcal{P}_{j}}
\vspace{-3mm}
\end{equation}
\end{footnotesize}
\subsection{Security model}
\label{Security_model}
The security model describes the level of data security achieved when transmitting the data from the UE to the O-RU. It depends on the data security level associated with selected encryption technique which depends on the  key size associated with it. In general, the larger the key size the less probable and the more difficult it is for an adversary to break the encryption \cite{4358705}. The robustness of encryption algorithms is estimated by evaluating the computational effort required to break the cipher through cryptanalysis. Ideally, a secure encryption system would require an exhaustive search with exponential complexity, making it impractical to break. However, practical systems may have vulnerabilities that allow for shortcut attacks, reducing the complexity needed to break the cipher. The expression for robustness often comes from the worst-case scenario, such as the brute force attack on the AES cipher, where the complexity is \(2^{128}\) for a 128-bit key. This motivates the use of a security measure like \(S(N) = \log_2(N)\), where \(N\) is the encryption key size. This measure reflects the difficulty of breaking the encryption based on key length, assuming no shortcut attacks are available.

Accordingly, the security level accomplished by UE $i \in \mathcal{I}$ at time step $t \in \mathcal{T}$ is determined with respect to brute force attacks as follows:
\begin{footnotesize}
\begin{equation}
    S_{i,t}= \log_2(N_{i}^{t}), \forall i \in \mathcal{I}, \forall t \in \mathcal{T}
\end{equation}
\end{footnotesize}
where $N_{i}^{t}$ is the key length used by UE $i$ at time step $t$.

\section{Problem Formulation} \label{sec:problem_formulation}

In this work, we aim to associate each UE $i \in \mathcal{I}$ with an O-RU $j \in \mathcal{J}$ at each time step $t \in \mathcal{T}$ to minimize latency and maximize security during data transmission. There is always a trade-off between these objectives as higher security levels increase latency due to the computational complexity of robust algorithms. This highlights the importance of optimizing the selection of encryption algorithm and its key length for each data exchange. Efficient resource allocation is needed, especially given limited device resources and varying O-RU security requirements. Some devices have low battery or limited computational capacity, restricting their association with high-security O-RUs. Additionally, the selection of the encryption algorithm to ensure a high security level for each UE will contribute to the latency overhead. Therefore, efficient resource utilization, security, and latency all need careful consideration when managing users association. 
The weight $\alpha$ represents the relative importance of security and latency depending on the application.

To formulate this joint problem, we define the Multi-objective Optimization Problem (MOP) with 3 decision variables. We model the association binary decision variable by $x_{i,j}^{t}$ defined as:
\begin{footnotesize}
\begin{equation}
    x_{i,j}^{t} = \left\{\begin{matrix}
\text{1, if UE $i \in \mathcal{I}$ is associated}\\
\text { with O-RU $j \in \mathcal{J}$ at time step $t \in \mathcal{T}$}\\
\text{0, otherwise} 
\end{matrix}\right.
\end{equation}
\end{footnotesize}

Additionally, we model the selection of an encryption algorithm $k$ by the binary decision variable $A_{i,k}^{t}$ defined as:
\begin{footnotesize}
\begin{equation}
    A_{i,k}^{t} = \left\{\begin{matrix}
\text{1, if encryption algorithm $k \in \mathcal{K}$ is selected}\\
\text{by UE $i \in \mathcal{I}$ at time step $t \in \mathcal{T}$}\\
\text{0, otherwise} 
\end{matrix}\right.
\end{equation}
\end{footnotesize}

Lastly, we model the key length applied by UE $i \in \mathcal{I}$ at time step $t \in \mathcal{T}$ with the variable $N_{i}^{t}$ where the key length must belong to the union of the possible options for all encryption algorithms ${\mathcal{D}_k}$ and it should be between the minimum and maximum key lengths of the chosen algorithm $k \in \mathcal{K}$, where:
\begin{footnotesize}
\begin{equation}
 \sum_{k=1}^K A_{i,k}^{t} D^k_{\text{min}} \leq N_{i}^{t} \leq \sum_{k=1}^K A_{i, k}^{t} D^k_{\text{max}}, \quad \forall i \in \mathcal{I}, \forall t \in \mathcal{T}
\end{equation}
\end{footnotesize}
\begin{footnotesize}
\begin{equation}
{{N}_{i}^{t}} \in \cup_{k=1}^{K}{\mathcal{D}_k}, \quad \forall i \in \mathcal{I}, \forall t \in \mathcal{T}
\end{equation}
\end{footnotesize}

From the discussion in Section \ref{sec:system_model}, the total latency to transmit data from UE $i$ to the selected O-RU $j$ at a time step $t$ is described as:
\begin{footnotesize}
\begin{equation}
    L_{i,t}= \sum_{k=1}^{K} A_{i,k}^{t} \cdot \tau_{enc}^{i,k,t} + \sum_{j=1}^{J} x_{i,j}^{t} (\tau_{comm}^{j,k,t} + \sum_{k=1}^{K}  A_{i,k}^{t} \cdot \tau_{dec}^{j,k,t}), \forall i \in \mathcal{I}, \forall t \in \mathcal{T}
\end{equation}
\end{footnotesize}

The problem formulation aiming at minimizing the latency and maximizing the security level relying on optimizing UEs association, encryption algorithms, and key lengths, is presented in (\ref{eq:P1})
Note that both objectives are normalized between 0-1 by dividing by their maximum values to ensure a fair trade-off between the objectives. The notation $[\cdot]^{\sim}$ represents the normalization function.
\begin{subequations}\label{eq:P1}
\footnotesize
\allowdisplaybreaks
\begin{align} 
 \textbf{P1: }  &\min_{\bm{x_{i,j}^{t},N_{i}^{t},A_{i,k}^{t}}} \quad (1 - \alpha) \sum_{t=1}^{T}{\sum_{i=1}^{I}{1-[S_{i,t}]^{\tilde{}}}} + \alpha \sum_{t=1}^{T}{\sum_{i=1}^{I}{[L_{i,t}]^{\tilde{}}}} \label{eq:P} \\
 \text{s.t.} \quad & \sum_{j=1}^{J}{x_{i,j}^{t} \cdot W_{j}} \leq S_{i,t} , \forall i \in \mathcal{I} \enspace \forall t \in \mathcal{T} \label{eq:Smin}\\
 &\sum_{j=1}^{J}{x_{i,j}^{t}} = 1, \forall i \in \mathcal{I} \enspace \forall t \in \mathcal{T} \label{eq:association} \\
 &\sum_{i=1}^{I}{x_{i,j}^{t}} \leq M_{j} , \forall j \in \mathcal{J} \enspace \forall t \in \mathcal{T} \label{eq:resource_blocks} \\
 &\sum_{k=1}^{K}{A_{i,k}^{t}} = 1, \forall i \in \mathcal{I} \enspace \forall t \in \mathcal{T} \label{eq:chosen_algo} \\
 &\sum_{k=1}^{K}{A_{i,k}^{t}} \cdot C_k \leq \Gamma_i, \forall i \in \mathcal{I}, \forall t \in \mathcal{T} \label{eq:computational_power} \\
  &\sum_{t=1}^{T}{[ (\eta_{cp}^{i,t}) + (\eta_{cm}^{i,t})]} \leq Z_i, \forall i \in \mathcal{I} \label{eq:power_consumption} \\
 &\sum_{k=1}^K A_{i,k}^{t} D^k_{\text{min}} \leq N_{i}^{t} \leq \sum_{k=1}^K A_{i, k}^{t} D^k_{\text{max}}, \forall i \in \mathcal{I}, \forall t \in \mathcal{T}\label{eq:n_value1} \\
 &{{N}_{i}^{t}} \in \cup_{k=1}^{K}{\mathcal{D}_k} \forall i \in \mathcal{I}, \forall t \in \mathcal{T} \label{eq:n_value2} \\
 &A_{i,k}^{t}, x_{i,j}^{t} \in \{0,1\}, \forall i \in \mathcal{I}, \forall j \in \mathcal{J}, 
 \forall k \in \mathcal{K}, \forall t \in \mathcal{T} \label{A_x_lim}
\end{align}
\end{subequations}
where $\eta_{cp}^{i,t}$ and $\eta_{cm}^{i,t}$ represent the power needed for the computations and communication respectively: 
\begin{footnotesize}
\begin{equation}
    \eta_{cp}^{i,t}= \sum_{k=1}^{K} A_{i,k}^{t} \cdot \tau_{enc}^{i,k,t} \cdot E_{cp}, \forall i \in \mathcal{I}, \forall t \in \mathcal{T}
\end{equation}
\end{footnotesize}
\begin{footnotesize}
\begin{equation}
    \eta_{cm}^{i,t}= \sum_{j=1}^{J} x_{i,j}^{t} \cdot \tau_{comm}^{i,j,t} \cdot E_{comm}, \forall i \in \mathcal{I}, \forall t \in \mathcal{T}
\end{equation}
\end{footnotesize}
where $E_{cp}$ and $E_{comm}$ represent the computation power consumption and transmission power consumption per second.

Constraint (\ref{eq:Smin}) ensures that the security level of every UE at all time steps is equal to or greater than the security requirement of the O-RU that the UE is connected to. Constraint (\ref{eq:association}) ensures that the UE is connected to only one O-RU at each time step. Constraint (\ref{eq:resource_blocks}) ensures that for all O-RUs at all time steps, the number of UEs connected to O-RU does not exceed the maximum available RBs for the O-RU where each UE is assigned to one RB at each time step. Constraint (\ref{eq:chosen_algo}) ensures that each UE selects only a single encryption algorithm at each time step. Constraints (\ref{eq:computational_power}) and (\ref{eq:power_consumption}) ensure that the allocation does not exceed the maximum computational and battery resources on UE. Additionally, constraint (\ref{A_x_lim}) ensures that both $x_{i,j}^{t}$ and $A_{i,k}^{t}$ are binary. 

\section{Solution approaches} \label{sec:solution}
This research addresses a Mixed Non-Linear Integer Program (MNLIP), an NP-hard problem due to variable multiplication in the objective function. We relaxed the problem and solved the relaxed version of the MOP in (\ref{eq:P1}) using two approaches to obtain sub-optimal solutions. First, we present the one-shot approach, solving the convexified and relaxed MOP P1 in (\ref{eq:P1}). Then, we introduce the iterative method to simplify and address the MOP complexities.
\subsection{One-shot Approach} \label{One-shot_approach}
In this approach, we start by relaxing the problem. In the relaxed version, we have omitted the decision variable ${N}_{i}^{t}$. This simplification is possible because, in our specific case, the key length, represented by ${N}_{i}^{t}$, can be derived directly from the binary decision variable ${A}_{i,k}^{t}$ through variables consolidation process. The selected encryption algorithms (DES, AES, RSA) support different key lengths, and there is no overlap among these lengths. Meaning, by selecting the key length, the algorithm is indirectly chosen. Therefore, the new variable $A_{i,g}^{t}$ is defined to indicate the selection of a specific key length $g$ for UE $i$ at time step $t$, where $g \leq |\cup_{k=1}^{K}{\mathcal{D}_k}|$, which is of size 8 in our study. This consolidation reduces the number of variables, focusing the optimization process on the remaining binary decision variables $A_{i,g}^{t}$ and $x_{i,j}^{t}$ and hence, simplifying the overall approach.

A common strategy to solve MNLIP problems involves relaxing the integer variables to formulate the problem as a Geometric Program (GP). However, this approach is not applicable in our case because of constraints (\ref{eq:association}) and (\ref{eq:chosen_algo}). In this paper, in order to convexify the problem, we propose to perform an exponential variable transformation for all the variables such that:
\begin{equation}
\lambda = \exp(\bar{\lambda})
\end{equation}
where $\lambda$ can be either $A_{i,g}^{t}$ or $x_{i,j}^{t}$. Such transformation is done to avoid the multiplications and make the MOP in (\ref{eq:P1}) as  exponential of the sum, which is known to be convex. Then, we relax all integer variables, namely, $A_{i,g}^{t}$ and ${x}_{i,j}^{t}$ by converting them from binary to continuous values in [0,1] which will be later floored after finding the solution of the simplified convex problem. Consequently, the MOP in (\ref{eq:P1}) can be reformulated as:

\begin{subequations}\label{eq:P2}
\footnotesize
\allowdisplaybreaks
\begin{align} 
 \textbf{P2: }  &\min (1 - \alpha) \sum_{t=1}^{T}{\sum_{i=1}^{I}{1 - [\bar{S}_{i,t}]^{\tilde{}}}} + \alpha \sum_{t=1}^{T}{\sum_{i=1}^{I}{[\bar{L}_{i,t}]^{\tilde{}}}} \label{eq:P_reform} \\
 \text{s.t.} \quad & (\ref{eq:Smin}), (\ref{eq:resource_blocks}), (\ref{eq:computational_power}), (\ref{eq:power_consumption})\\ 
 & \sum_{j=1}^{J}\exp({\bar{x}_{i,j}^{t}}) - 1 \leq 0, \forall i \in \mathcal{I} \enspace \forall t \in \mathcal{T} \label{eq:association_leq} \\
  & 1 - \sum_{j=1}^{J}\exp({\bar{x}_{i,j}^{t}}) \leq 0, \forall i \in \mathcal{I} \enspace \forall t \in \mathcal{T} \label{eq:association_geq} \\
 & \sum_{m=1}^{\mathcal{J}-1} \sum_{n=m+1}^{\mathcal{J}} \exp({x_{i,m}^{t}} + {x_{i,n}^{t}}) \leq \epsilon, \forall i \in \mathcal{I}, \forall t \in \mathcal{T} \label{eq:mult_x} \\
 & \sum_{k=1}^{|\cup_{k=1}^{K}{\mathcal{D}_k}|} \exp({\bar{A}_{i,k}^{t}}) - 1 \leq 0, \forall i \in \mathcal{I} \enspace \forall t \in \mathcal{T} \label{eq:chosen_algo_leq} \\
  & 1 - \sum_{k=1}^{|\cup_{k=1}^{K}{\mathcal{D}_k}|} \exp({\bar{A}_{i,k}^{t}}) \leq 0, \forall i \in \mathcal{I} \enspace \forall t \in \mathcal{T} \label{eq:chosen_algo_geq} \\
  & \sum_{m=1}^{|\cup_{k=1}^{K}{\mathcal{D}_k}|-1} \sum_{n=m+1}^{|\cup_{k=1}^{K}{\mathcal{D}_k}|} \exp({A_{i,m}^{t}} + {A_{i,n}^{t}}) \leq \epsilon, \forall i \in \mathcal{I}, \forall t \in \mathcal{T} \label{eq:mult_A} \\
 & \bar{x}_{i,j}^{t}, \bar{A}_{i,k}^{t} \in \Psi \label{A_x_lims} 
\end{align}
\end{subequations}
where $\Psi$ is the new domain of the MOP
after the reformulation and $\epsilon$ is a very small number, and $\epsilon > 0$.
Moreover, we applied exponential variable
transformation on the expressions of $S_{i,t}$, $L_{i,t}$, $\eta_{cp}^{i,t}$, and $\eta_{cm}^{i,t}$. Constraints (\ref{eq:association}) and (\ref{eq:chosen_algo}) are converted to (\ref{eq:association_leq}), (\ref{eq:association_geq}), (\ref{eq:chosen_algo_leq}, and (\ref{eq:chosen_algo_geq}). Moreover, we add constraint (\ref{eq:mult_x}) to ensure that each UE can only associate with one O-RU where its association variable has higher value, and enforce the values of other O-RUs' association variables to be close to 0. Similarly, constraint (\ref{eq:mult_A}) ensures that each UE choose only one encryption algorithm.
\subsection{Iterative approach} \label{Iterative_approach}
The second approach aims to optimize values through an iterative process by breaking the initial problem into smaller sub-problems and iteratively optimizing individual variables. In this approach we have omitted the decision variable ${N}_{i}^{t}$, and the variables are not relaxed to continuous values. The method fixes one variable and solves for the other, with the output of the first problem serving as the input to the next. Starting with initial values for $x_{i,j}^{t}$, it alternates between solving for $A_{i,k}^{t}$ and $x_{i,j}^{t}$ using the objective function $F_u$ at each iteration $u$ and the corresponding constraints until it converges either by reaching the predefined maximum number of iterations $u_{max}$ or if the absolute difference between the objective function values of consecutive iterations is less than a specified tolerance $\epsilon$. The details of the iterative approach are provided in Algorithm \ref{alg:iterative_approach_algo}.

{\footnotesize
\begin{algorithm} 
\caption{Iterative Approach}
\label{alg:iterative_approach_algo}
\KwIn{$\epsilon$,  $u_{max}$}
\KwOut{Optimized values $A^*$, $x^*$}
Initialize $x \gets x_0$\;
Initialize $u \gets 1$\;
\Repeat{$|F_u - F_{u-1}| < \epsilon$ or $u>u_{max}$}{ 
    Solve for $A$: 
    \Indp $A^*_u \gets F_u(x^*_{u-1})$\; 
    s.t. (\ref{eq:Smin}), (\ref{eq:chosen_algo}), (\ref{eq:computational_power}), (\ref{eq:power_consumption}), (\ref{A_x_lim}) 
    \Indm Solve for $x$: 
    \Indp $x^*_u \gets F_u(A^*_u)$\; 
    s.t. (\ref{eq:Smin}), (\ref{eq:association}), (\ref{eq:resource_blocks}), (\ref{eq:power_consumption}), (\ref{A_x_lim}) 
    $u \gets u + 1$\;
    \Indm Compute objective function value $F_{u}$\;
}
\end{algorithm}
}

\section{Simulation Results} \label{sec:results}
This section presents numerical results to evaluate our proposed approaches and compare them to the optimal solution derived through exhaustive search. Simulations were run with 3 O-RUs and 4 UEs over 3 time steps, using parameters: $M_j=3$, $W_j$ from [6, 12] which are determined using the logarithm of the key length, $\rho_{i,j}^{t}$ from [10, 100] bps, $PT_{i,t}$ from [50, $2\times10^4$] KB, $\Gamma_i$ from [656, $1.7\times10^7$], $E_{cp}=4W$, $E_{comm}=7W$, ${\mathcal{Q}_{i}}$ from [1.8, 2.4] GHz, ${\mathcal{P}_{j}}$ from [3.5, 3.9] GHz, and $Z_i$ from [460, $2\times10^6$]J.All parameters are uniformly selected from the given ranges. It is worth noting that the one-shot approach is classified as Signomial Program (SP), which can be solved using GPkit \cite{Burnell2020GPkitAH} while we used CVX to solve the iterative approach.

\begin{figure*}[!t]
    \centering
    \mbox{
        \begin{subfigure}[b]{0.23\textwidth}
            \centering
            \includegraphics[scale=0.4]{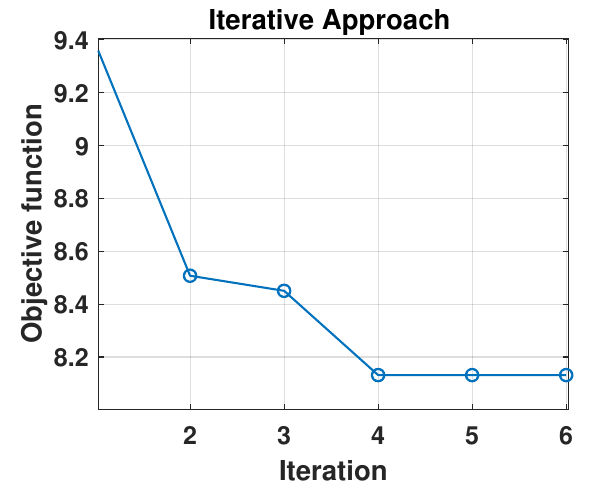}
            \caption{}
            \label{iterative_convergence}
        \end{subfigure}
        \hspace{-2.5mm}
        \begin{subfigure}[b]{0.23\textwidth}
            \centering
            \includegraphics[scale=0.4]{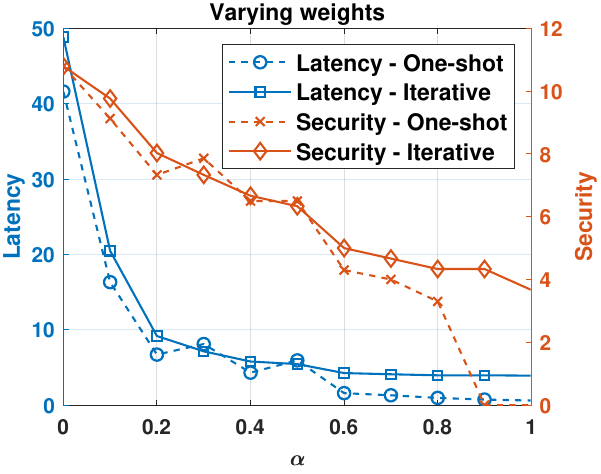}
            \caption{}
            \label{varying_a}
        \end{subfigure}
        \hspace{-2.5mm}
        \begin{subfigure}[b]{0.23\textwidth}
            \centering
            \includegraphics[scale=0.4]{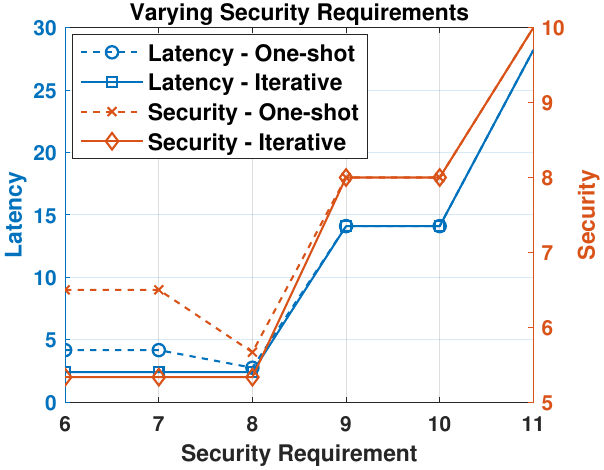}
            \caption{}
            \label{varying_w}
        \end{subfigure}
        \hspace{-2.5mm}
        \begin{subfigure}[b]{0.23\textwidth}
            \centering
            \includegraphics[scale=0.4]{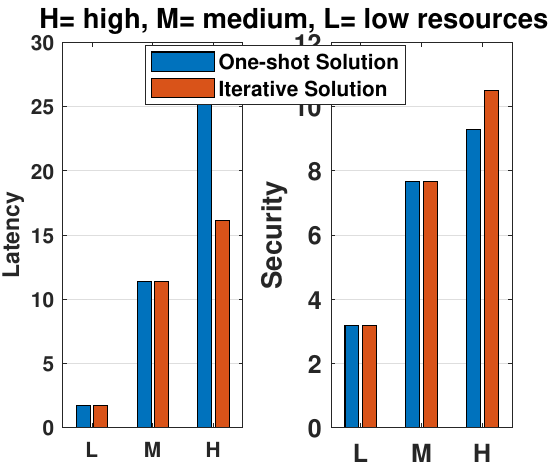}
            \caption{}
            \label{varying_r}
        \end{subfigure}
    }
    \vspace{-2mm}
    \mbox{
        \begin{subfigure}[b]{0.23\textwidth}
            \centering
            \includegraphics[scale=0.4]{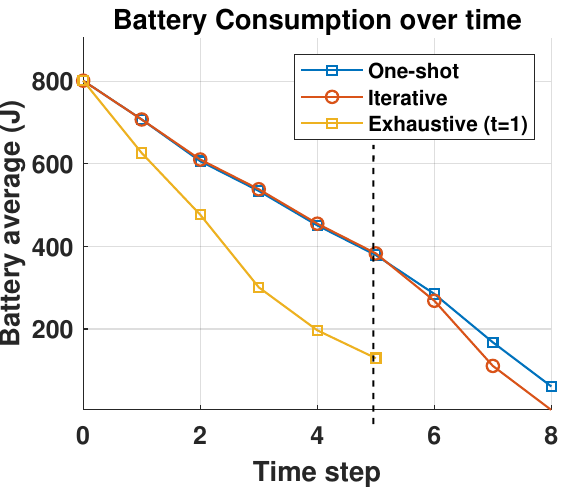}
            \caption{}
            \label{opt_t}
        \end{subfigure}
        \hspace{-2.5mm}
        \begin{subfigure}[b]{0.23\textwidth}
            \centering
            \includegraphics[scale=0.4]{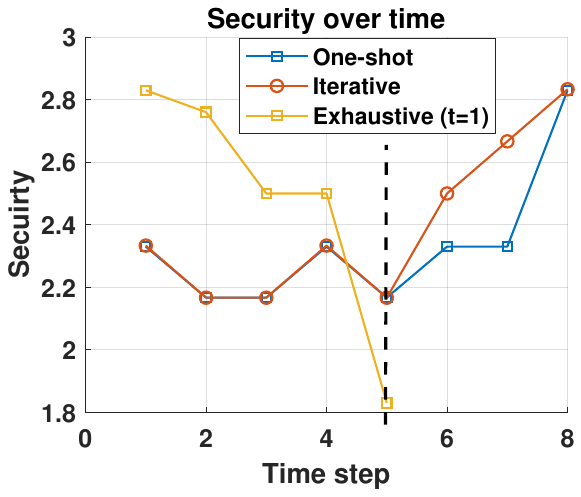}
            \caption{}
            \label{Bi_sec}
        \end{subfigure}
        \hspace{-2.5mm}
        \begin{subfigure}[b]{0.23\textwidth}
            \centering
            \includegraphics[scale=0.4]{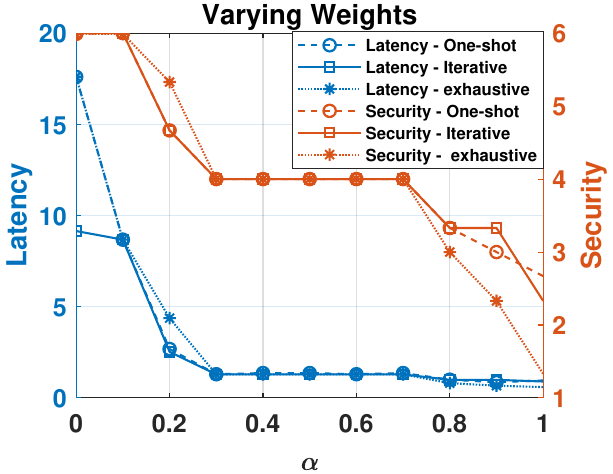}
            \caption{}
            \label{varying_a_ex}
        \end{subfigure}
        \hspace{-2.5mm}
        
        \begin{subfigure}[b]{0.23\textwidth}
            \centering
            \includegraphics[scale=0.4]{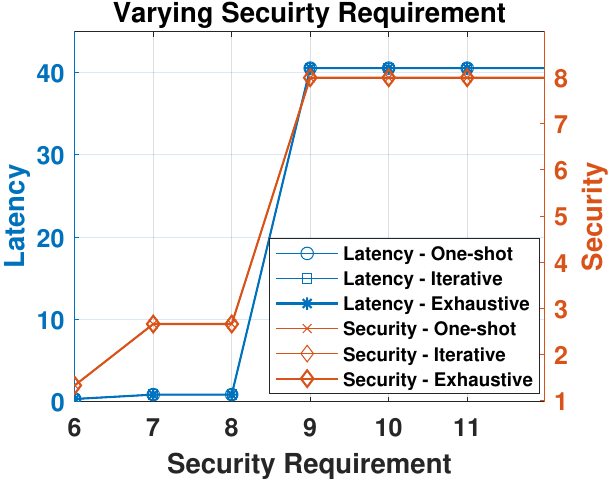}
            \caption{}
            \label{varying_w_ex}
        \end{subfigure}
    }
    \vspace{2mm}
    \caption{Performance of our approaches in terms of latency, security, and battery level}
    \label{performance}
    \vspace{-3mm}
\end{figure*}

Figure (\ref{iterative_convergence}) shows the convergence behavior of Algorithm \ref{alg:iterative_approach_algo} adopted in the iterative approach over a series of iterations. It shows that the objective function consistently decreased and the gradient norms approached zero, indicating convergence to a local minimum, and hence a sub-optimal solution.

Figure (\ref{varying_a}) shows the trade-off between normalized latency and normalized security for iterative and one-shot solutions as $\alpha$ varies. Both approaches accomplish the expected trade-off between latency and security. Latency decreases with higher $\alpha$, prioritizing latency, while security improves when it is the focus. The one-shot approach achieves the lowest latency with less robust encryption, whereas the iterative approach maintains higher security at the cost of increased latency. When security is prioritized, both approaches perform similarly, with the iterative approach slightly better. Overall, the one-shot solution performs better with a holistic view, though it is more complex. Solving times were 6 minutes for the one-shot and 30 seconds for the iterative approach, tested on an Intel i9-13900K, 64GB RAM, Windows 11 system.

Figure (\ref{varying_w}) illustrates the normalized latency and normalized security when varying O-RU security requirements with $\alpha$=0.5. 
High security demands force devices to use robust algorithms, which significantly increasing the latency. The problem was infeasible for a security requirement of 12, requiring all devices to use RSA with the highest key value. With low security requirements, the one-shot approach achieves better security with comparable latency, slightly outperforming the iterative approach.
It can be noticed that the one-shot solution dropped at 8. This is because the decision variables are continuous instead of being binary which makes the results not stable which is not the case with the iterative approach. 

Figure (\ref{varying_r}) shows the normalized latency and normalized security when varying the resources of the devices. We simulated three scenarios where all devices have either low, medium, or high computational and battery resources. In this experiment, $\alpha$=0.1, which means security is given higher priority. It can be noticed that higher resources result in higher security and hence, higher latency. This is because devices with more resources can afford to use more robust encryption algorithms, which provide better security but also require more computational power and time, leading to increased latency.

In figure (\ref{opt_t}) we studied the battery consumption. Our approaches solve the problem over an extended period $\mathcal{T}$, considering future resource allocation. Exhaustive search for high $\mathcal{T}$ values is complex, so we applied it at each step $t$ separately, updating battery consumption based on completed encryption tasks. Initially, exhaustive search achieves higher security as shown in figure (\ref{Bi_sec}). However, after four steps, the battery depletes, as the exhaustive search does not account for future encryption tasks, failing to balance security with the need to preserve battery life for upcoming operations. Although it is faster to solve each time step $t$ separately, it does not give optimal solutions for period $\mathcal{T}$ and in most of the cases it cannot preserve battery life for the whole duration. This justify opting for relaxed solutions at the cost of slightly lower performance.

Lastly, figures (\ref{varying_a_ex}) and (\ref{varying_w_ex}) compare the performance of both approaches to exhaustive search in terms of security and latency over 2 time steps and 3 key lengths (64, 256, 4096 bits) due to the complexity of exhaustive search. Figure (\ref{varying_a_ex}) shows exhaustive search performs better, but other approaches provide near-optimal solutions. In figure (\ref{varying_w_ex}), all approaches yield the same optimal solution. Our solutions are near-optimal with lower complexity, justifying the relaxed methods. Solving times were: exhaustive search 7 minutes, one-shot 4 minutes, iterative 20 seconds. The results align with previous observations on varying alpha and security requirements, aiming to compare our results to optimal ones under different configurations.
\vspace{-2mm}
\section{Conclusion} \label{sec:conclusion}
In this work, we addressed the problem of efficient resource management in O-RAN. We formulated a MOP for UE-O-RU association, encryption algorithm selection, and resource management, aiming to minimize total latency and maximize security. Various approaches were employed to solve this problem. Simulation results indicate that our proposed methods are close to the optimal solution, demonstrating their effectiveness. However, due to the centralized and computationally expensive nature of the optimization approach, these solutions are not efficient for larger setups. Future work will focus on developing a solution more suitable for online setups.
\vspace{-2mm}
\section{ACKNOWLEDGEMENT} \label{sec:acknowledgement}
This work was jointly supported by Qatar University and the University of Guelph - IRCC-2023-171. The findings achieved herein are solely the responsibility of the authors.

\end{document}